%% file: main.tex
\documentclass[9pt, sigconf,bookmarks=false,screen]{acmart}
\AtBeginDocument{%
  }
    
\input{packages}
\input{macro}
\graphicspath{{./figs/}}
\usepackage{footnote}

\usepackage{geometry}
\usepackage{tabularx,booktabs}
\newcolumntype{L}[1]{>{\raggedright\arraybackslash}p{#1}}
\newcolumntype{C}[1]{>{\centering\arraybackslash}p{#1}}
\newcolumntype{R}[1]{>{\raggedleft\arraybackslash}p{#1}}
\newcolumntype{Y}{>{\centering\arraybackslash}X}

\copyrightyear{2025} 
\acmYear{2025} 
\setcopyright{acmcopyright}\acmConference[ICCAD '25]{IEEE/ACM International Conference on Computer-Aided Design}{October 26--30, 2020}{Munich, Germany}
\acmBooktitle{IEEE/ACM International Conference on Computer-Aided Design (ICCAD '25), October 26--30, 2025, Munich, Germany}
\acmPrice{15.00}
\acmISBN{979-8-4007-1293-7/25/03}

\begin{document}

\expandafter\def\expandafter\normalsize\expandafter{%
    \normalsize%
    \setlength\abovedisplayskip{0pt}%
    \setlength\belowdisplayskip{8pt}%
    \setlength\abovedisplayshortskip{-8pt}%
    \setlength\belowdisplayshortskip{2pt}%
}

\settopmatter{printacmref=false} %
\renewcommand\footnotetextcopyrightpermission[1]{} %
\pagestyle{plain} %

\newcommand*{\affaddr}[1]{#1} %
\newcommand*{\affmark}[1][*]{\textsuperscript{#1}}

\title{
OptiClear: Differentiable Curvilinear Design Rule Legalization for Inverse-Designed Photonic Devices
}

\author{
Hongjian Zhou\textsuperscript{1},
Haoyu Yang\textsuperscript{3},
Nicholas Gangi\textsuperscript{2},
Zhaoran (Rena) Huang\textsuperscript{2},
Jiaqi Gu\textsuperscript{1,$\dagger$}\\
\textsuperscript{1}Arizona State University \quad
\textsuperscript{2}Rensselaer Polytechnic Institute \quad
\textsuperscript{3}NVIDIA Corporation \\
\textsuperscript{$\dagger$}jiaqigu@asu.edu
}

\input{doc/0_abstract}
\maketitle
\input{doc/1_intro}

\input{doc/2_prelim}
\input{doc/3_algo}

\input{doc/4_result}

\input{doc/5_conclu}

{
\balance
\footnotesize

\input{main.bbl}
}

\end{document}

%% file: packages.tex
\usepackage[utf8]{inputenc} %
\usepackage[T1]{fontenc}    %

\usepackage[normalem]{ulem}
\usepackage{pifont}
\usepackage{xcolor}         %
\usepackage{booktabs}
\usepackage{hyperref}
\hypersetup{colorlinks,
    linkcolor=red,citecolor=citecolor,urlcolor=blue}
\usepackage{algorithm}
\usepackage{graphicx}
\usepackage{threeparttable}
\usepackage{multirow}
\usepackage{amsmath}
\usepackage{bm}
\usepackage{bbm}
\usepackage{amsfonts}
\usepackage{amsthm}
\usepackage[]{algpseudocode}
\usepackage{enumitem} %
\usepackage[subrefformat=parens,labelformat=parens]{subfig}
\usepackage{colortbl}

\usepackage{wrapfig}

\setcitestyle{sort&compress,numbers}
\usepackage{xspace}

%% file: macro.tex
\definecolor{citecolor}{RGB}{34,139,34}
\definecolor{mydarkblue}{rgb}{0,0.08,1}
\definecolor{mydarkgreen}{rgb}{0.02,0.6,0.02}
\definecolor{mydarkred}{rgb}{0.8,0.02,0.02}
\definecolor{mydarkorange}{rgb}{0.40,0.2,0.02}
\definecolor{mypurple}{RGB}{111,0,255}
\definecolor{myred}{rgb}{1.0,0.0,0.0}
\definecolor{mygold}{rgb}{0.75,0.6,0.12}
\definecolor{myblue}{rgb}{0,0.2,0.8}
\definecolor{mydarkgray}{rgb}{0.,0.2,0.2}

\definecolor{lightred}{RGB}{255,235,235}
\definecolor{lightgreen}{RGB}{235,255,235}
\definecolor{lightblue}{RGB}{235,235,255}
\definecolor{lightcyan}{RGB}{235,255,255}
\definecolor{lightmagenta}{RGB}{255,235,255}
\definecolor{lightyellow}{RGB}{255,255,235}

\definecolor{qxkcolor}{RGB}{215,235,255}
\definecolor{softmaxcolor}{RGB}{230,235,255}
\definecolor{probxvcolor}{RGB}{255,255,235}

\definecolor{topkcolor}{RGB}{255,235,235}
\definecolor{zecolor}{RGB}{255,255,235}
\definecolor{dynacolor}{RGB}{235,255,255}

\definecolor{reviewcolor}{RGB}{0,0,200}

\theoremstyle{plain}

\theoremstyle{definition}

\algdef{SE}[DOWHILE]{Do}{doWhile}{\algorithmicdo}[1]{\algorithmicwhile\ #1}%

\newcommand{\squishlist}{
 \begin{list}{$\bullet$}
  { \setlength{\itemsep}{0pt}
     \setlength{\parsep}{3pt}
     \setlength{\topsep}{3pt}
     \setlength{\partopsep}{0pt}
     \setlength{\leftmargin}{1.5em}
     \setlength{\labelwidth}{1em}
     \setlength{\labelsep}{0.5em} } }

\newcommand{\squishend}{
  \end{list}  }

\newcommand{\name}{\texttt{OptiClear}\xspace}
\newcommand{\nameR}{\texttt{OptiClear-R}\xspace}
\newcommand{\nameD}{\texttt{OptiClear-D}\xspace}

%% file: doc/0_abstract.tex
\begin{abstract}
\label{abstract}
Photonic inverse design enables ultra-compact, high-performance devices with highly curvilinear and non-intuitive geometries, but the resulting layouts often violate fabrication design rules and hinder foundry manufacturing.
Legalization methods designed for rectilinear Manhattan electrical layouts are not directly applicable to curvilinear inverse-designed photonic devices. Meanwhile, existing fabrication-aware inverse-design methods apply soft penalties on small features and sharp curvatures, but still cannot guarantee design-rule-compliant final layouts.
In this work, we present \name, a curvilinear design rule legalization framework for inverse-designed photonic devices. 
\name provides two complementary legalization engines: \nameR, a rule-based morphological legalizer that efficiently resolves violation regions through iterative morphology-guided mask processing, and \nameD, a differentiable legalizer that formulates legalization as a minimum-distortion mask optimization problem under morphological stationary-point constraints, explicitly seeking a rule-compliant layout with minimal geometric deviation from the original design.
We further develop customized differentiable morphological GPU operators that significantly improve the scalability of high-resolution mask legalization.
Comprehensive evaluation across diverse inverse-designed photonic devices and a wide range of design-rule settings shows that \name reduces design-rule violations from thousands to zero. 
The rule-based legalizer offers high runtime efficiency, while the differentiable legalizer more faithfully preserves the original optical functionality.
This work establishes curvilinear design rule legalization as a practical post-design electronic-photonic design automation (EPDA) stage for translating high-performance inverse-designed photonic layouts into manufacturable tape-out-ready devices.
The differentiable formulation also opens the door to future end-to-end design-rule-aware curvilinear inverse hardware design.
\end{abstract}

%% file: doc/1_intro.tex
\section{Introduction}
\label{sec:Introduction}

Photonic integrated circuits (PICs) are increasingly important for emerging applications such as optical interconnects, lidar, biosensing, quantum photonics, and optical computing, due to their advantages in speed, bandwidth, and energy efficiency~\cite{zhu2026harnessingphotonicsmachineintelligence,NP_JLT2024_Ning,NP_NanoPhotonics2020_Feng, NP_Nature2025_Hua, NP_Nature2025_Ahmed, NP_NaturePhotonics2021_Shastri}.
As PIC complexity continues to grow, there is increasing demand for compact and high-performance photonic devices with advanced application-specific functionality.
In this context, photonic inverse design~\cite{NP_minkov2020inverse} has emerged as a powerful paradigm for device synthesis. 
By optimizing the material permittivity distribution over a much larger design space than conventional parameterized design, inverse design has enabled ultra-compact photonic components with highly non-intuitive geometries and excellent optical performance on diverse applications.

However, the layouts produced by photonic inverse design are often extremely challenging to manufacture in foundry flows~\cite{wang2019robust}.
Unlike conventional hand-designed photonic devices with relatively regular geometry, inverse-designed devices typically exhibit highly curvilinear, non-Manhattan boundaries, dense subwavelength details, irregular notches, and narrow gaps. As a result, the final layouts frequently violate fabrication design rules, including minimum width, minimum spacing, area, curvature, and related curvilinear mask constraints.
A non-compliant layout may require a design-rule waiver (e.g., AIM Photonics), may be rejected by the foundry (e.g., GlobalFoundries, TSMC), and carry high risks of pattern distortion, yield loss, and degraded functionality after fabrication. 
Therefore, a practical path from high-performance inverse-designed masks to manufacturable tape-out-ready layouts is critically needed, but \emph{so far does not exist in current electronic-photonic design automation (EPDA) flow}.

A central challenge is that legalization for inverse-designed photonic layouts is fundamentally different from legalization for rectilinear electrical masks.
Inverse-designed photonic devices typically exhibit fine features, irregular boundaries, and highly curvilinear geometries, in which width, spacing, curvature, and area constraints are tightly \emph{coupled and often mutually conflicting}. 
Fixing one violation can therefore easily create new ones, making legalization a \emph{discrete and highly nonconvex process rather than a sequence of simple local edits}. 
This is fundamentally more difficult than existing legalization and mask-repair methods in electrical physical design, which rely on Manhattan geometry and local operations such as edge pushing and shape snapping.
These assumptions are poorly suited to inverse-designed photonic layouts, which are strongly curvilinear, topologically complex, and highly sensitive to small geometric perturbations~\cite{park2026interpretablegeometrysensitivityinverse, zhou2026prismphotonicsinformedinverselithography}. 
Directly adapting Manhattan-style DRV fixing to inverse-designed photonic layouts is often ineffective: it does not naturally handle curvilinear violation patterns, and it provides little control over the geometry distortion.
As a result, effective photonic legalization must \textbf{jointly enforce manufacturability and preserve geometry similarity} to the original high-performance mask, since even small distortions can significantly degrade device functionality.

Despite this clear need, current photonic inverse-design flows still lack a dedicated legalization stage. 
Most prior inverse design research optimizes primarily for optical performance, while fabrication compliance is handled only approximately through manufacturability regularization or left unresolved.
Some fabrication-aware inverse-design (FAID) methods~\cite{NP_chen2020design, mao2023multi, Hammond2021Photonic} attempt to improve manufacturability by introducing soft penalties on minimum feature size or post-processing, e.g., gaps/curvature penalty~\cite{DFM_vercruysse2019analytical}, filtered density fields, or post-design open/close fixing~\cite{Hammond2021Photonic}.
However, these methods remain soft manufacturability priors rather than explicit legalizers, and thus cannot guarantee rule-clean final layouts. 
This leaves a missing post-design legalization stage for curvilinear inverse-designed photonic devices.

To close this gap, we \textbf{define and enable, for the first time, a dedicated curvilinear design-rule legalization stage for inverse-designed photonic devices}. 
We present \name, a morphology-based curvilinear mask legalization framework that explicitly targets manufacturable curvilinear layouts with minimal deviation from the original design. 
The framework includes two complementary legalizers. 
\nameR is a rule-based morphological legalizer that efficiently resolves violation regions through iterative morphology-guided mask updates. 
\nameD is a differentiable legalizer that formulates legalization as a minimum-distortion mask optimization problem under morphological stationary-point constraints, solved using an augmented Lagrangian framework with customized differentiable morphological operators. 
In this way, \name provides both an efficient rule-based path and a higher-fidelity differentiable path with runtime-quality trade-offs for curvilinear photonic legalization.

The main contributions of this work are summarized as follows:
\begin{itemize}
    \item We define and enable, for the first time, an explicit curvilinear design-rule legalization stage for inverse-designed photonic devices, filling a critical gap in current photonic inverse-design and EPDA flows.
    \item We propose a morphology-based legalization flow \name with two complementary engines: a rule-based legalizer for efficient conflict resolution and a differentiable legalizer that minimizes mask distortion under morphological stationary-point constraints via an augmented Lagrangian method.
    \item We develop customized differentiable morphological GPU operators with 450$\times$ faster runtime and 280$\times$ lower memory consumption than state-of-the-art third-party library, making scalable legalization of high-resolution masks practical.
    \item We demonstrate across diverse inverse-designed photonic devices that \name reduces design-rule violations from thousands to zero in minutes while minimizing optical functionality degradation, remains effective across a broad range of spacing and width rules, and complements fabrication-aware inverse design in producing manufacturable, DRV-clean layouts.
\end{itemize}

%% file: doc/2_prelim.tex
\vspace{-5pt}
\section{Preliminaries}
\label{sec:Preliminaries}

\subsection{Photonic Curvilinear Design Rules}

Photonic layouts are subject to several types of fabrication design rules to ensure manufacturability and yield~\cite{chrostowski2015silicon}. Typical rules include minimum width, minimum spacing, minimum area, notch, enclosure, and port alignment constraints. Representative examples of these violations are illustrated in Fig.~\ref{fig:mot_drv}.

Compared with conventional electrical IC layouts, photonic layouts are fundamentally different in geometry. Electrical layouts are typically dominated by Manhattan structures composed of horizontal and vertical edges, for which rule checking is relatively straightforward. In contrast, photonic devices often contain curved boundaries, oblique edges, and highly irregular freeform geometries, especially in inverse-designed structures. As a result, photonic layouts are more sensitive to local curvature, narrow necks, and small-angle boundary interactions. These effects are less critical in standard Manhattan layouts but become important in photonic devices, where geometric distortion can directly affect both manufacturability and optical performance.

For curvy photonic structures, curvature-related manufacturability is not always checked explicitly in existing foundry PDKs. Instead, most PDKs approximate curvature constraints through geometric distance-based rules. In practice, this is usually reduced to width and spacing checks between polygon edges. For example, in the open-source silicon photonics PDK SiEPiC~\cite{SiEPIC_PDK}, when the included angle between polygon edges is smaller than $80^\circ$, an additional edge-to-edge distance check is applied. In this way, by enforcing the minimum width and minimum spacing rules under small-angle conditions, the DRC indirectly constrains the local curvature of the layout.

\input{figtex/fig_mot_drv}
\subsection{Rectilinear Legalization in Electronics}
Prior work in electronic design automation (EDA) has addressed geometry repair and rule-aware layout generation mainly in rectilinear, grid-constrained layout domains, especially for standard-cell and routing applications.
Representative efforts incorporate DRC awareness during layout synthesis or routing, for example through reinforcement-learning-based routing, routability-aware placement, or iterative geometry refinement to reduce width, spacing, via, and cut-metal violations~\cite{PD_ren2021nvcell,PD_ho2023nvcell,PD_chung2024optimal,PD_liang2021generating,PD_huang2025drc}. 
However, these methods are fundamentally tailored to Manhattan electrical layouts, where violations are typically repaired through routing adjustment, coordinate refinement, edge movement, or local polygon optimization on rectilinear geometries. 
Such assumptions do not transfer directly to inverse-designed photonic devices, whose masks are strongly curvilinear, topologically complex, and highly sensitive to small geometric perturbations. 
To the best of our knowledge, \uline{explicit post-design legalization for curvilinear inverse-designed layouts has not been previously studied}.

\subsection{Fabrication-aware Inverse Design (FAID)}
Most prior efforts on manufacturable photonic inverse design improve fabrication robustness during optimization without explicit design rule legalization. 
A first class of methods incorporates analytical fabrication constraints, such as spatial derivative-approximated gap and curvature penalties, into levelset-based inverse design to suppress non-manufacturable features during optimization~\cite{vercruysse2019analytical}. 
A second class uses density- or filtering-based regularization, including Gaussian blurring, low-pass filtering, and related smoothing operators, to discourage tiny features and narrow gaps in the generated layouts~\cite{NP_gershnabel2022reparameterization,hammond2022high,NP_khoram2020controlling, chen2025bi}. 
A third class integrates lithography or process-variation models into the optimization loop to improve robustness against fabrication-induced deformation~\cite{boson}.
While these methods improve manufacturability, they act primarily through \emph{soft regularization} or robustness objectives during design optimization. 
They do not explicitly enforce the rigorous design rules on the discretized and polygonized layout, and therefore cannot guarantee a DRV-clean result. 
Therefore, fabrication-aware inverse design and explicit post-design legalization remain \uline{distinct and complementary capabilities to our legalization}.

\subsection{Morphological Operation}
Morphological operations provide a natural way to analyze and modify local geometric structures~\cite{raid2014image}. 
In this work, the device layout is represented as a binary/grayscale image, where the foreground denotes the material region, and the background denotes the surrounding cladding. 
Given a mask $M$ and a structuring element (kernel), morphology probes the local geometric support and separation of each pixel by sliding the kernel over the mask.

\input{figtex/fig_prelim_morphology}

Among the basic morphological operators, erosion shrinks the foreground by removing boundary pixels, while dilation expands the foreground by adding pixels around the boundary. Based on these two primitives, \emph{opening} and \emph{closing} are defined as
\begin{equation}
\small
\texttt{Open}(M) = \texttt{Dilate}(\texttt{Erode}(M)), \,
\texttt{Close}(M) = \texttt{Erode}(\texttt{Dilate}(M)).
\end{equation}

Opening removes small foreground protrusions, thin bridges, and isolated islands, while closing fills small holes, narrow gaps, and concave notches as shown in Fig~\ref{fig:openclose}. 
Therefore, opening is suitable for fixing small-width violations, whereas closing is effective for fixing small-spacing and hole-related violations.

%% file: figtex/fig_mot_drv.tex
\begin{figure}
    \centering
    \includegraphics[width=0.99\columnwidth]{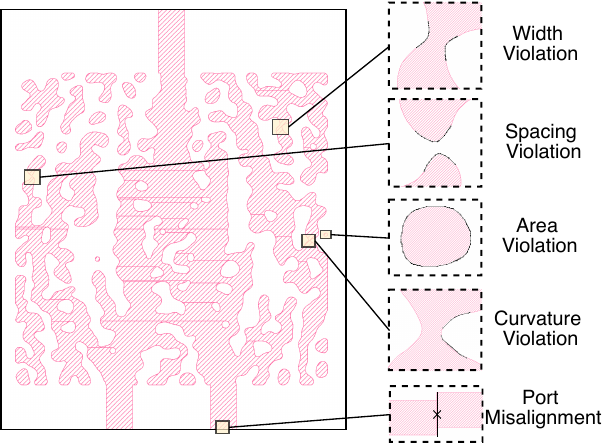}
    \vspace{-5pt}
    \caption{Example inverse-designed photonic device with representative fabrication design rule violations, which hinder the practical deployment of ID devices in manufacturable PICs.}
    \label{fig:mot_drv}
     \vspace{-5pt}
\end{figure}

%% file: figtex/fig_prelim_morphology.tex
\begin{figure}
    \centering
    \includegraphics[width=1\columnwidth]{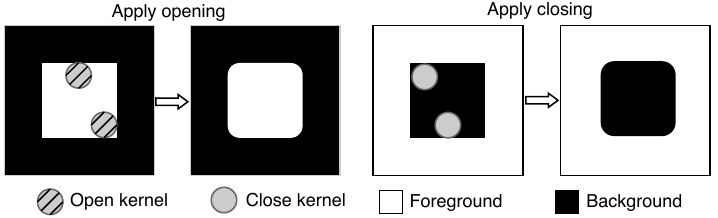}
    \vspace{-10pt}
    \caption{Morphological \texttt{Open} and \texttt{Close} operations.}
    \label{fig:openclose}
\end{figure}

%% file: doc/3_algo.tex
\section{Proposed \name Legalization Framework}

Curvilinear design rule legalization for inverse-designed photonic layouts is challenging because multiple fabrication constraints, such as minimum width, spacing, and curvature, are tightly coupled, while the device functionality is highly sensitive to geometric perturbations. 
Unlike Manhattan electrical layouts, these violations cannot be repaired reliably through local edge pushing or simple polygon adjustment. 
Instead, legalization must simultaneously enforce manufacturability and limit geometric distortion to the original high-performance design.

Our key observation is that many curvilinear DRVs can be interpreted by geometric survivability under morphological operations. 
As discussed in Section~\ref{sec:Preliminaries}, opening naturally removes insufficiently supported foreground features, while closing eliminates narrow gaps and enclosed voids. 
These properties make morphology a natural foundation for curvilinear rule legalization directly in the pixelated mask domain, which is naturally compatible with level-set-based inverse design and avoids fragile local edits in the polygon domain. 
However, naive opening and closing with a kernel size chosen only from the minimum-width rule are insufficient, since the two operations may conflict and cannot fully account for curvature-related constraints.

Based on this insight, we develop a morphology-based legalization framework with two complementary engines. 
The first is a rule-based legalizer, \nameR, that efficiently resolves morphology-induced conflict regions through iterative mask updates. 
The second is a differentiable legalizer, \nameD, that formulates legalization as minimum-distortion mask recovery under morphological stationary-point constraints.
A central component of both legalizers is the morphological operator. 
We first show an analytical derivation of the kernel size from the design rules and polygonization error, and then present the proposed legalization methods.

\input{figtex/fig_kernel_size}

\subsection{Morphological Kernel Size Derivation}
We first answer two central questions in morphology-based curvilinear legalization: \ding{202}~why processing in the pixelate mask domain and \ding{203}~how to choose the structuring-element (i.e., kernel) size so that mask-domain operations faithfully enforce the target curvilinear design rules after polygonization.

\subsubsection{\ding{202}~Why processing in the pixelated mask domain?}
Inverse-designed photonic devices are naturally represented as pixelated masks, where the foreground denotes silicon and the background denotes cladding materials such as silicon dioxide or air. 
For fabrication, the mask is converted into a polygonal GDS layout, typically through contour extraction followed by polygon simplification. 
Since the original mask boundary is pixelated and not perfectly smooth, nearby \emph{contour vertices are often merged during simplification}, introducing geometric discrepancy between the mask-domain representation and the final polygonal layout. 
In our framework, legalization is performed directly on the original pixel-domain mask, and polygon conversion is applied only once after legalization. 
This \textbf{avoids accumulating additional polygonization cost and error during intermediate processing}.

\subsubsection{\ding{203}~How to set morphological kernel size?}
\label{sec:kernel_size}
Choosing the kernel size is critical, since it determines which geometric features can survive morphology-based legalization. 
A natural starting point is the target minimum-width and minimum-spacing rules: the kernel radius must be larger than those rules to remove sub-rule foreground features and narrow gaps, yet not so large that it introduces excessive geometric distortion.
However, these \emph{nominal rules alone are insufficient for curvilinear photonic layouts}, where additional curvature violations can emerge after mask-to-polygon conversion, especially near small-angle corners.

In Fig.~\ref{fig:kernel_size}, let $\theta$ denote the included angle between two polygon edges. 
In rectilinear masks, edges are parallel or orthogonal, and minimum-width/spacing checks are governed by parallel (face-to-face) edge distance. 
In contrast, curvilinear layouts contain edge pairs with arbitrary relative angles. 
In curvilinear masks, width/spacing checks are activated only when the angle between two edges is below a threshold (e.g., $80^\circ$).
As $\theta$ decreases, the edge-to-edge separation shrinks geometrically, so a layout may violate the rule even when its nominal width or spacing appears sufficient. 
Therefore, the morphological kernel size may need to be larger than the nominal minimum-width/spacing threshold in order to satisfy this angle-dependent distance constraint. 
Let $EF$ denote the critical edge-to-edge distance required by the design rule. From the circle geometry in Fig.~\ref{fig:kernel_size}, the required \uline{target kernel size $K$}, i.e., diameter $BD$, can be derived as
\begin{equation}
\label{eq:kernel_size}
    K=BD = \frac{EF + 2EC \sin(\theta/2)}{\cos(\theta/2)},
\end{equation}
where $EC$ is a safety margin accounting for polygon simplification error. 
Thus, the \textbf{kernel size must account not only for nominal width and spacing rules, but also for corner-angle amplification from curvature rules and polygonization error}.

\subsection{\nameR: Rule-based Legalization}
Morphological opening and closing can effectively fix many simple DRVs. 
Hence, we are motivated to develop a heuristic rule-based legalizer to efficiently fix the violations via morphological operations.

\input{figtex/fig_conflict_zoom}
The \emph{key challenge}, however, is that naive application of opening and closing does not necessarily converge to a legal mask, because the two operations can \emph{conflict near marginal geometries}. 
A typical failure case is illustrated in Fig.~\ref{fig:conflict_zone}, which shows the \emph{conflict zone} between opening and closing. When two features with sizes comparable to the kernel are placed too close to each other, opening tends to erode the weak boundary regions and partially remove them. 
At the same time, because the spacing between the features is smaller than the closing kernel size, closing regrows material in the gap and forms a thin neck between them. 
This thin neck is then removed again by the next opening operation.
\textbf{Repeated open/close operations therefore fail to reach a \uline{morphological stationary point} and instead oscillate between incompatible configurations}.

This observation suggests that \uline{legalization is achieved when opening and closing reach an equilibrium}, i.e., no conflict zone remains. Formally, the legalized mask should satisfy the stationary condition,
\begin{equation}
    M = \texttt{Open}(M) = \texttt{Close}(M),
\end{equation}
where $M$ denotes the legalized mask. 
Under this stationary condition, applying either operation no longer changes the layout, indicating that unstable small features, narrow gaps, and hole-related conflicts have been resolved. 
Therefore, the key objective of the rule-based legalizer \nameR is to explicitly break conflict zones and project the mask toward this morphological stationary point.

To resolve such conflict zones, the most direct strategy is to handle them explicitly according to geometric rules. In essence, a conflict zone can be eliminated in two ways: either by increasing the spacing between nearby features so that the closing kernel can pass through without reconnecting them, or by directly merging the features so that the remaining structure can fully survive the opening kernel. Based on this idea, the overall flow of the proposed rule-based legalizer \nameR is illustrated in Fig.~\ref{fig:rule_flow}.

\input{figtex/fig_rule_flow}
We first apply \emph{small-kernel preprocessing open/close operations} to remove trivial islands and holes without over-modifying the main device geometry before handling the more difficult conflict regions. 
Specifically, the small kernel size is approximately half the target kernel size $K$ minus a safety margin. 
We then process the remaining foreground and background conflict regions separately.

For the foreground, we first extract the skeleton of the foreground region via \emph{skeletonization} operations and then dilate pixels along the skeleton using the target kernel. In this way, thin bridges or weak connections inside the foreground can be thickened, so that they will no longer be removed by the subsequent opening operation.
Then, for the background, we extract the skeleton and perform background dilation along it to widen narrow gaps and disconnect necks, thereby breaking the conflict zones caused by insufficient spacing. 
After each update, opening and closing are reapplied to suppress newly introduced artifacts and move the mask toward a more stable state.

Because foreground and background updates \emph{compete for the same physical area}, modifying one side may affect the legality of the other. 
To improve stability and better preserve the optically active device region, \nameR prioritizes the high-refractive-index foreground (i.e., the silicon region), because distortions in the high-index core are more likely to degrade mode confinement and introduce unwanted scattering, reflection, or radiation.
Therefore, after the initial background update, the subsequent iterations only continue foreground dilation until the convergence condition is satisfied. 
Once this stationary condition is reached, the mask is regarded as a legalized solution.

\subsection{\nameD: Differentiable Legalization}
As discussed in Section~\ref{sec:kernel_size}, the morphological kernel size must be chosen \emph{conservatively} to satisfy the target curvilinear design rules. 
In practice, this derived kernel size $K$ may \emph{overestimate} the needed width or spacing correction, and directly applying morphology with such a kernel can introduce unnecessary geometry distortion.
Since inverse-designed photonic devices are highly sensitive to geometric perturbations, effective legalization should \uline{explicitly minimize deviation from the original high-performance layout}.

To better preserve the original device geometry, we further propose the second engine, \nameD, a differentiable morphological legalizer that formulates legalization as a \textbf{minimum-distortion optimization problem under morphological stationary-point constraints}, as shown in Fig.~\ref{fig:differentiable_flow}.
\nameD searches for a legalized mask that remains close to the original design while satisfying the same morphology-based legality conditions, shown in Fig.~\ref{fig:SolutionSpace}.

Accordingly, we formulate legalization as the following constrained optimization problem:
\begin{equation}
\begin{aligned}
    &\quad\quad\min_{\beta} \quad  \|M(\beta) - M_{0}\|_2^2 \\
    \text{s.t.} \quad & \texttt{Open}(M(\beta))=M(\beta), \, \texttt{Close}(M(\beta))=M(\beta),\\
    &M(\beta) = \big[\texttt{Tanh}\big(\eta \cdot \phi(\beta)\big)+1\big]/2
\end{aligned}
\label{eq:legalization_problem}
\end{equation}
where $M_{0}$ denotes the original mask and $M$ denotes the mask parametrized by a leveset function $\phi$ that is controlled by latent variables $\beta$, and $\eta$ control the sharpness of the binarization to preserve gradients, which is set to 20 here.
The objective explicitly minimizes geometric distortion relative to the original design, while the constraints require the optimized mask to be invariant under both opening and closing. 

To solve this constrained problem efficiently, we further reformulate it using the augmented Lagrangian method (ALM) as
\begin{equation}
\begin{aligned}
\mathcal{L}(M,\lambda,\rho)
=&\ \|M-M_{0}\|_2^2 + \lambda \, \bigl(\texttt{Open}(M)-\texttt{Close}(M)\bigr) \\
&+ \rho \, \bigl\|\texttt{Open}(M)-\texttt{Close}(M)\bigr\|_2^2.
\end{aligned}
\label{eq:alm_legalization}
\end{equation}
where $\lambda$ denotes the Lagrange multiplier and $\rho$ is the penalty coefficient. 
Note that $\lambda$ is not a scalar but a \emph{mask-shaped multiplier field} with the same spatial dimensions as $M$, so that the constraint can be enforced in a pixel-wise manner. 
The second and third terms penalize the inconsistency between the opened and closed masks, driving the solution toward the morphological stationary legal mask. 
In this way, legalization becomes a differentiable optimization process that jointly enforces manufacturability and minimizes geometry distortion.

After each optimization iteration, the multiplier field is updated using the current residual between the opened and closed masks:
\begin{equation}
\lambda^{(t+1)}
=
\lambda^{(t)}
+
\rho \, \bigl(\mathrm{Open}(M^{(t)})-\mathrm{Close}(M^{(t)})\bigr),
\label{eq:lambda_update}
\end{equation}
where $t$ denotes the iteration step. 
The Lagrange multipliers are updated to gradually enforce the stationarity constraint, with the local update magnitude determined \emph{adaptively} by the degree of inconsistency between $\mathrm{Open}(M)$ and $\mathrm{Close}(M)$ at each pixel.
The optimization dynamics are shown in the curve in Fig.~\ref{fig:SolutionSpace}.

\input{figtex/fig_opt_flow}
\input{figtex/fig_solution_space}
\subsection{Triton-accelerated Efficient Differentiable Morphological Operators}
\label{sec:triton_morph}

\input{figtex/fig_triton_kornia}
Morphological operators are a central component of both the rule-based and optimization-based legalizers, particularly in the latter where differentiable open/close operators must be evaluated repeatedly within the optimization loop. 
When applied to high-resolution design masks with large structuring elements, these operators become a major computational bottleneck. 
Existing differentiable implementations in third-party libraries such as Kornia are built on dense neighborhood expansion followed by a large 2D convolution with channel-wise min/max reduction. 
This formulation scales poorly because it materializes a high-dimensional intermediate tensor whose size grows with both image resolution and kernel area. 
For a $4096 \times 4096$ mask and a $101 \times 101$ structuring element, Kornia's implementation involves a huge 2D convolution $(1, 1, 4096, 4096) * (101^2, 1, 101, 101)\rightarrow (1, 101^2, 4096, 4096)$ with a channel-wise max/min operation (along the $101^2$ dimension), which requires prohibitive 637 GB memory allocation for the intermediate tensors, making such implementations impractical for large-scale differentiable DRV fixing.

To overcome this limitation, we develop a custom Triton-based differentiable morphological operators that reformulate the computation around \textbf{sparse active-kernel traversal} and \textbf{fused local reduction}. 
Instead of explicitly expanding each pixel into its local neighborhood, the proposed operator directly traverses only the active offsets of the structuring element and performs the corresponding min/max reduction on the fly on GPU. 
For grayscale morphology, the structuring-element weights are fused into the same traversal. 
This design changes the complexity bottleneck from dense neighborhood materialization to \emph{direct reduction over the active kernel support}, substantially reducing memory traffic and improving scalability for large kernels.

\input{figtex/fig_benchmark}
The proposed operator is also designed to preserve exact hard morphological semantics during gradient-based optimization. 
The forward pass stores only the winning offset index for each output pixel, sufficient for exact hard min/max backpropagation while avoiding the memory overhead of saving expanded neighborhood values.  
Based on these optimized erosion and dilation primitives, opening and closing are implemented compositionally with the same efficiency benefits.

The core contribution is not only a faster implementation but a \emph{more scalable differentiable morphology formulation for large-area inverse-designed layout processing}. 
As shown in Fig.~\ref{fig:TritonKorniaComparison}, the proposed Triton operator maintains nearly constant peak memory and achieves over \textbf{453$\times$ speedup} at moderate kernel sizes, while \textbf{avoiding the out-of-memory} behavior even with large kernels. 
This capability is a key enabler for practical, large-scale differentiable DRV fixing.

%% file: figtex/fig_kernel_size.tex
\begin{figure}
    \centering
    \includegraphics[width=0.85\columnwidth]{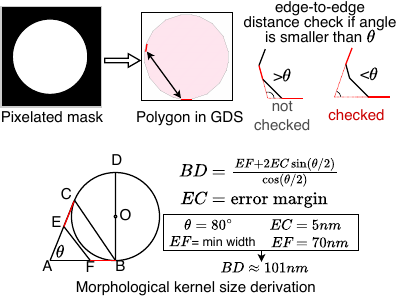}
    \vspace{-5pt}
    \caption{Illustration of design rule check for curvilinear photonic device masks the derivation of the morphological kernel size. 
    DRVs are fixed directly on the pixelated mask by morphological operations, and the legalized mask is then converted into GDS polygons for final design rule checking. 
    Morphological kernel size is derived by considering the edge-to-edge distance rule, curvature requirement, and the error margin.}
    \label{fig:kernel_size}
\end{figure}

%% file: figtex/fig_conflict_zoom.tex
\begin{figure}
    \centering
    \includegraphics[width=1\columnwidth]{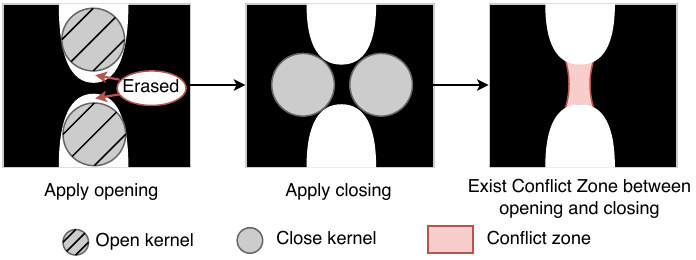}
    \caption{Conflict zone between morphological opening and closing. Opening removes narrow foreground connections, while closing tends to reconnect nearby features across small gaps. The effects of opening and closing are inconsistent, making it difficult to completely eliminate all design rule violations.}
    \label{fig:conflict_zone}
\end{figure}

%% file: figtex/fig_rule_flow.tex
\begin{figure}
    \centering
    \includegraphics[width=1\columnwidth]{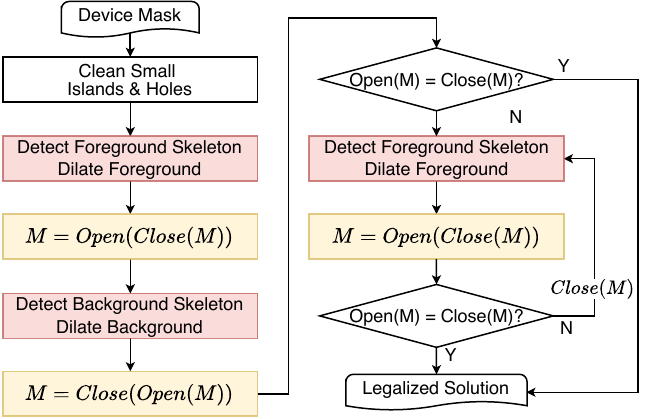}
    \vspace{-10pt}
    \caption{Flow of the proposed rule-based legalizer \nameR. 
    Foreground dilation is prioritized to preserve the silicon region, and the procedure iterates until the opened mask and closed mask become consistent, yielding a legalized solution.}
    \label{fig:rule_flow}
\end{figure}

%% file: figtex/fig_opt_flow.tex
\begin{figure}
    \centering
    \includegraphics[width=1\columnwidth]{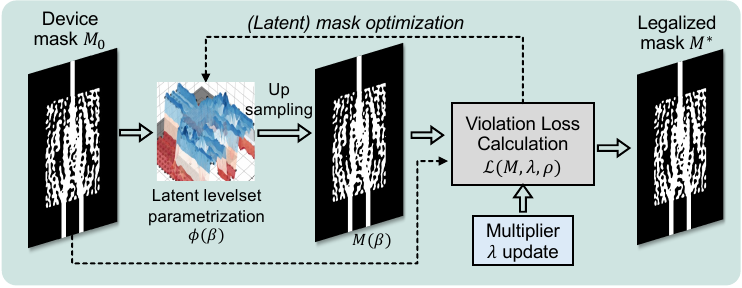}
    \vspace{-10pt}
    \caption{Flow of the proposed differentiable legalizer \nameD. 
    The mask is optimized in the latent space via a level-set parameterization initialized from the original mask.}
    \label{fig:differentiable_flow}
\end{figure}

%% file: figtex/fig_solution_space.tex
\begin{figure}
    \centering
    \includegraphics[width=0.99\linewidth]{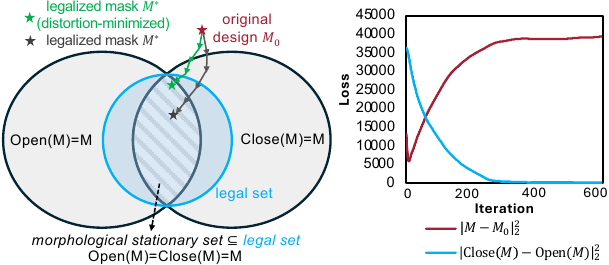}
    \vspace{-5pt}
    \caption{Solution space illustration and optimization dynamics of \nameD.}
    \label{fig:SolutionSpace}
\end{figure}

%% file: figtex/fig_triton_kornia.tex
\begin{figure}
    \centering
    \includegraphics[width=0.99\columnwidth]{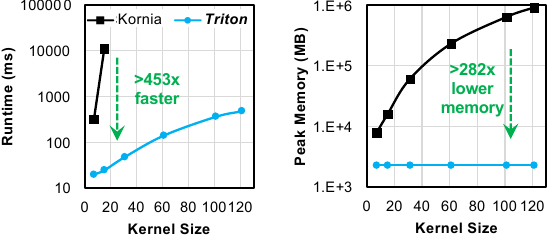}
    \vspace{-5pt}
    \caption{On 4096$\times$4096 masks with various \emph{Opening} kernel sizes, our Triton-based kernel inference runs orders-of-magnitude faster without OOM issues.}
    \label{fig:TritonKorniaComparison}
\end{figure}

%% file: figtex/fig_benchmark.tex
\begin{figure*}
    \centering
    \includegraphics[width=0.99\textwidth]{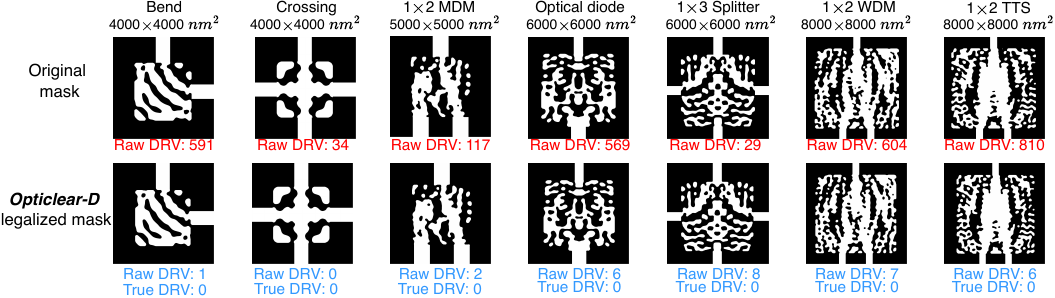}
    \vspace{-10pt}
    \caption{Benchmark examples of inverse-designed photonic devices. The original device masks contain a large number of design rule violations. 
    After \nameD legalization, the \uline{remaining raw DRVs are all caused by false-positive DRC detections from KLayout and can be filtered out (thus True DRV=0) (see Section~\ref{sec:false_drv})}.}
    \label{fig:benchmark}
\end{figure*}

%% file: doc/4_result.tex
\section{Evaluation Results}
\label{sec:ExperimentalResults}

We evaluate \name\ from four aspects: (1) its ability to eliminate curvilinear design-rule violations (DRVs) on diverse inverse-designed photonic devices, (2) the tradeoff between the efficient rule-based \nameR\ and the higher-fidelity differentiable \nameD, (3) its robustness under different spacing and width rule settings, and (4) its compatibility with fabrication-aware inverse design.

\subsection{Benchmarks, Baselines, and Metrics}
The proposed \name framework is developed in PyTorch on a Linux server equipped with a 64-core AMD EPYC 7763 CPU (base frequency 2.45\,GHz) and four NVIDIA RTX A6000 GPUs, each with 48\,GB of VRAM.
We use a \uline{mask resolution of 1 nm/pixel} in all methods.
For \nameD, we optimize 600 steps with Adam optimizer with a learning rate of 0.1.
The $\rho$ in ALM is set to 0.2. 
Levelset sharpness $\beta$=20.
We use the same opening and closing kernel size in both \nameD and \nameR, which is derived according to~\eqref{eq:kernel_size}.

\noindent\textbf{Inverse-designed Device Benchmarks}.~
We benchmark \name\ on representative inverse-designed photonic devices generated by the open-source framework~\cite{MAPS}, covering diverse functionalities and layout sizes ranging from $4000\times4000~\mathrm{nm}^2$ to $8000\times8000~\mathrm{nm}^2$, as shown in Fig.~\ref{fig:benchmark}. During inverse design, a soft Gaussian blurring penalty is applied to control the minimum feature size (MFS)~\cite{MAPS}. 
For the examples in Fig.~\ref{fig:benchmark}, the blur kernel size is set to 80\,nm. Nevertheless, even with this fabrication-aware MFS control, the resulting masks still exhibit a large number of DRVs under the 70\,nm minimum-spacing and minimum-width rules. 
This confirms that soft manufacturability regularization alone is insufficient to guarantee DRV-clean layouts, thereby necessitating our explicit post-design legalization.

\noindent\textbf{Baseline}.~
As a baseline, we adopt the manufacturability-aware optimization strategy in~\cite{vercruysse2019analytical}, which introduces soft gap and curvature penalties on the latent levelset to discourage rule-violating geometries.
We adapt this idea as a \emph{Gap\&Curvature Penalty} baseline and use it to assess whether soft manufacturability penalties alone can serve as an alternative to explicit legalization.

\noindent\textbf{Evaluation Metrics}.~
We evaluate legalization quality from three complementary perspectives: \uline{geometry preservation, optical functionality, and final design-rule compliance}. 
Since legalization is performed directly in the pixelated mask domain, we first measure geometric distortion relative to the original design using the $\ell_2$ distance and intersection-over-union (IoU):
\begin{equation}
D_{\ell_2} = \|M - M_0\|_2^2, \qquad
\mathrm{IoU}(M,M_0)=\frac{|M\cap M_0|}{|M\cup M_0|},
\end{equation}
where $M_0$ and $M$ denote the original and legalized masks, respectively. 
Smaller $D_{\ell_2}$ and larger IoU indicate better geometry preservation.
We then re-simulate the legalized masks using rigorous full-wave finite-difference frequency-domain (FDFD) simulation and report the post-legalization \uline{figure of merit (FoM), i.e., S-parameters}, which reflects how well the original optical functionality is preserved. 
To evaluate final manufacturability, the legalized masks are converted to GDS and design rule checking (DRC) is performed in KLayout.
The process design kit used in our experiments is the open-source silicon photonics PDK SiEPiC~\cite{SiEPIC_PDK}, where the default minimum-spacing and minimum-width rules for silicon are 70\,nm for edge pairs with relative angle smaller than $80^\circ$. 
We \emph{mainly focus on width and spacing rules}, which are primary sources of DRVs.
Minimum-area, enclosure, and other layer-interaction rules are outside the current scope.
Finally, we also report the \uline{runtime (RT)} of the legalization process on a single A6000 GPU to evaluate its computational efficiency. 

\subsection{False-Positive Design Rule Violation Filtering}
\label{sec:false_drv}
\input{figtex/fig_drv_filter}
During evaluation, we observed that a small number of reported violations persist even when the corresponding original polygon-edge pairs should not trigger rule checking because they do not satisfy the rule-angle threshold, as illustrated in Fig.~\ref{fig:drv_filter}. 
These reports arise from internal polygon simplification in the KLayout DRC engine: during rule checking, simplified or merged edges may be used in place of the original polygon edges, which can introduce discrepancies in angle and edge-to-edge distance and lead to false-positive violation reports. 
Raw DRV is revalidated on the original polygon geometry, and is discarded only if the original edge pair fails the angle-trigger condition or satisfies the minimum spacing/width rule.
We therefore filter out false-positive violations and only report true DRVs.

\subsection{Main Result}

\input{tables/tab_exp_main}

The main result in this section shows that explicit legalization is both necessary and effective for inverse-designed photonic devices: both \nameR and \nameD \textbf{drive \#DRV to zero across diverse benchmarks}, while \nameD better preserves optical functionality and \nameR achieves higher runtime efficiency.

Table~\ref{tab:main_result} compares \name with open/close-based legalization~\cite{Hammond2021Photonic} the Gap/Curvature-Penalty baselines~\cite{vercruysse2019analytical}.
Both \nameR\ and \nameD\ \uline{fix all true DRVs on evaluated inverse-designed devices}, \textbf{with only a few (<8) remaining false-positive DRVs that are filtered out}.
The soft gap/curvature-penalty baseline fails to eliminate violations in most cases.
Although the open/close baseline can achieve small L2 distances and high IoUs, it still leaves a large number of DRVs.
Therefore, it is meaningless to compare their post-fixing FoM with ours.
This confirms that fabrication-aware regularization alone is insufficient to guarantee rule-clean layouts.

The second key observation is that the two proposed legalizers offer complementary benefits.
Both preserve the post-legalization figure of merit (FoM) well relative to the original masks, but \nameD consistently achieves better functionality preservation, especially on larger-footprint and more performance-sensitive devices such as WDM and TTS. This difference is illustrated in Fig.~\ref{fig:fixing_comparison}.
For the spacing violation, the baseline fails to remove it. 
In contrast, \nameD resolves it by introducing the minimum spacing correction, whereas \nameR repairs it by merging nearby silicon regions to preserve the foreground. 
Because this region lies close to the input port and is highly performance-sensitive, the distortion-minimized optimization in \nameD leads to better FoM preservation.

A further \uline{insight} from these results is that \textbf{global geometry similarity metrics, such as $L_2$ distance and IoU, do not accurately predict post-legalization optical performance}. 
Inverse-designed photonic devices are highly sensitive to local perturbations in critical regions, so a method with better global shape similarity may still exhibit worse FoM if it perturbs a functionally important area. 
This observation further motivates distortion-aware legalization rather than purely rule-driven mask repair.

Finally, the runtime results imply an efficiency-fidelity tradeoff between two legalizers. 
\nameR provides faster legalization by using rule-based morphology updates, whereas \nameD incurs higher runtime due to iterative differentiable optimization on high-resolution masks. 
This additional cost, however, enables \nameD to better control geometry distortion and preserve optical functionality after legalization.

\input{figtex/fig_fixing_comparison}

\subsection{Ablation Study}
\input{figtex/fig_ablation_mse}
\subsubsection{Effect of Distortion Loss}
In Fig.~\ref{fig:MSE_compare}, we further ablate the role of the distortion-minimization term in \nameD by removing $|M-M_0|_2^2$ and optimizing only the ALM objective. 
Although this variant can still enforce legality, it produces legalized masks with larger deviations from the original design. 
This confirms that distortion minimization is critical, as it suppresses unnecessary geometry changes and steers the optimization toward performance-preserving legal solutions.

\input{tables/tab_exp_mfs}

\input{tables/tab_exp_design_rule}

\subsubsection{Synergy with Fabrication-Aware Inverse Design}
We first study how fabrication-aware inverse design (FAID) interacts with the proposed legalization by varying the minimum feature size (MFS) control during inverse design.
This study examines how upstream manufacturability regularization affects the legalization quality, and it demonstrates the capability of \name across diverse devices under different design rule requirements ranging from 70\,nm to 130\,nm.

Table~\ref{tab:mfs_result} reports the results for devices designed with different MFS controls, while the target legalization rules are fixed to the default 70\,nm spacing/width rules in SiEPiC~\cite{SiEPIC_PDK}. As the MFS increases, the inverse-design space becomes more constrained, and the original device FoM generally decreases. At the same time, the performance gap between the original and legalized layouts becomes smaller, indicating that \emph{stronger MFS control reduces the amount of geometric correction required during legalization}.
This trend reveals that \textbf{FAID and legalization play distinct but synergistic roles.}
Stronger MFS regularization can reduce the legalization burden by suppressing fine and fragile features during inverse design, but it still does not eliminate the need for explicit post-design legalization. 
In particular, under the most aggressive setting with MFS = 70\,nm, \nameD shows noticeably better FoM preservation than \nameR, indicating that \textbf{distortion-aware legalization becomes especially important when the original layout contains finer and more performance-sensitive structures}. 
As the MFS increases, the post-legalization performance of \nameR and \nameD becomes closer, since the layouts become easier to legalize with less geometric and functional perturbation needed.

\subsubsection{Evaluation under Different Design Rules}
We evaluate \name under different spacing and width rules using devices generated with MFS = 110\,nm, in Table~\ref{tab:design_rule_result}. 
Overall, \name remains effective to generate DRV-clean layout across a broad range of rule settings. The post-legalization FoM degrades as the target design rules become more stringent.
This trend highlights that \emph{legalization and inverse design are complementary and should be co-optimized}.
When the target width/spacing rules become significantly tighter than the MFS, the legalizer must introduce larger geometric corrections, which in turn causes greater functionality degradation. 
A \uline{general guideline} is that the manufacturability setting used during inverse design, such as MFS control, should be chosen with awareness of the final fabrication rules to achieve better post-legalization performance.

The runtime results also reveal different scaling behaviors for the two legalizers. 
Stricter design rules require larger morphological kernels. 
\uline{Thanks to the accelerated morphology operators, the runtime of \nameD is relatively insensitive to kernel size} and is dominated mainly by device size. 
In contrast, the runtime of \nameR increases more noticeably with rule size, since it involves CPU-based skeleton extraction and iterative conflict-zone processing.

\subsubsection{Mask Resolution}
Table~\ref{tab:exp_resolution} shows that full-resolution masks are necessary to ensure legalization quality, while downsampled masks fail to generate DRV-clean masks, which also justifies the value of our customized Triton operators.
\input{tables/tab_exp_resolution}

%% file: figtex/fig_drv_filter.tex
\begin{figure}
    \centering
    \includegraphics[width=0.87\columnwidth]{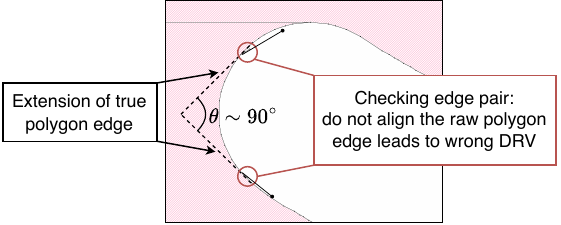}
    \vspace{-5pt}
    \caption{False-positive DRV reported by KLayout due to internal polygon edge merging, which causes incorrect edge-pair checking after polygon simplification, thus needs to be ignored.}
    \label{fig:drv_filter}
     \vspace{-5pt}
\end{figure}

%% file: tables/tab_exp_main.tex
\begin{table*}[t]
\caption{Legalization results on inverse-designed photonic devices with an 80\,nm Gaussian-blur-based minimum feature size control under 70\,nm minimum spacing and width rules. The compared metrics include the number of true design rule violations (\#DRV), figure of merit (FoM), $L_2$ loss, intersection-over-union (IoU), and runtime (RT) (s).}
\vspace{-8pt}
\resizebox{0.999\textwidth}{!}{
\begin{tabular}{|c|cc|ccccc|ccccc|ccccc|ccccc|}
\hline
& \multicolumn{2}{c|}{Design mask}
& \multicolumn{5}{c|}{Morphological Open/Close~\cite{Hammond2021Photonic}}
& \multicolumn{5}{c|}{Gap\&Curvature-based Penalty~\cite{vercruysse2019analytical}}
& \multicolumn{5}{c|}{\nameR}
& \multicolumn{5}{c|}{\nameD} \\
\multirow{-2}{*}{Device}
& \#DRV & FoM
& \#DRV & FoM & L2 & IoU & RT
& \#DRV & FoM & L2 & IoU & RT
& \#DRV & FoM & L2 & IoU & RT
& \#DRV & FoM & L2 & IoU & RT \\
\hline

Bending
& {\color[HTML]{C00000}586} & 0.970
& {\color[HTML]{C00000}532} & {\color[HTML]{808080}0.970} & 37246 & 0.998 & 14
& {\color[HTML]{C00000}344} & {\color[HTML]{808080}0.970} & 50668 & 0.989 & 44
& \textbf{0} & 0.969 & 38110 & 0.995 & 64
& \textbf{0} & 0.970 & 37659 & 0.990 & 565 \\
\hline

Crossing
& {\color[HTML]{C00000}28} & 0.936
& \textbf{0} & 0.935 & 17735 & 0.999 & 22
& \textbf{0} & {\color[HTML]{808080}0.936} & 47703 & 0.993 & 44
& \textbf{0} & 0.936 & 17860 & 0.999 & 93
& \textbf{0} & 0.936 & 17427 & 0.996 & 568 \\
\hline

& & 0.997
& & 0.994 & & & 
& & 0.996 & & & 
& & 0.996 & & &
& & 0.996 & & & \\
\multirow{-2}{*}{\begin{tabular}[c]{@{}c@{}}Optical\\diode\end{tabular}}
& \multirow{-2}{*}{{\color[HTML]{C00000}562}}
& 4.8E-5
& \multirow{-2}{*}{{\color[HTML]{C00000}513}}
& 8.7E-6
& \multirow{-2}{*}{48210}
& \multirow{-2}{*}{0.997}
& \multirow{-2}{*}{32}
& \multirow{-2}{*}{{\color[HTML]{C00000}340}}
& {\color[HTML]{808080}4.0E-4}
& \multirow{-2}{*}{93085}
& \multirow{-2}{*}{0.985}
& \multirow{-2}{*}{85}
& \multirow{-2}{*}{\textbf{0}}
& 3.9E-5
& \multirow{-2}{*}{47952}
& \multirow{-2}{*}{0.997}
& \multirow{-2}{*}{240}
& \multirow{-2}{*}{\textbf{0}}
& 3.2E-5
& \multirow{-2}{*}{47961}
& \multirow{-2}{*}{0.991}
& \multirow{-2}{*}{1271} \\
\hline

& & 0.327
& & 0.327 & & & 
& & 0.326 & & & 
& & 0.317 & & &
& & 0.318 & & & \\
& & 0.330
& & 0.330 & & & 
& & 0.328 & & & 
& & 0.334 & & &
& & 0.333 & & & \\
\multirow{-3}{*}{\begin{tabular}[c]{@{}c@{}}1$\times$3\\Splitter\end{tabular}}
& \multirow{-3}{*}{{\color[HTML]{C00000}21}}
& 0.330
& \multirow{-3}{*}{\textbf{0}}
& 0.330
& \multirow{-3}{*}{42308}
& \multirow{-3}{*}{0.998}
& \multirow{-3}{*}{37}
& \multirow{-3}{*}{{\color[HTML]{C00000}15}}
& {\color[HTML]{808080}0.329}
& \multirow{-3}{*}{98658}
& \multirow{-3}{*}{0.984}
& \multirow{-3}{*}{84}
& \multirow{-3}{*}{\textbf{0}}
& 0.334
& \multirow{-3}{*}{43891}
& \multirow{-3}{*}{0.997}
& \multirow{-3}{*}{236}
& \multirow{-3}{*}{\textbf{0}}
& 0.334
& \multirow{-3}{*}{41955}
& \multirow{-3}{*}{0.991}
& \multirow{-3}{*}{1266} \\
\hline

& & 0.943
& & 0.923 & & & 
& & 0.942 & & & 
& & 0.916 & & &
& & 0.918 & & & \\
\multirow{-2}{*}{MDM}
& \multirow{-2}{*}{{\color[HTML]{C00000}108}}
& 0.948
& \multirow{-2}{*}{{\color[HTML]{C00000}100}}
& 0.892
& \multirow{-2}{*}{53017}
& \multirow{-2}{*}{0.995}
& \multirow{-2}{*}{22}
& \multirow{-2}{*}{{\color[HTML]{C00000}81}}
& {\color[HTML]{808080}0.948}
& \multirow{-2}{*}{69466}
& \multirow{-2}{*}{0.991}
& \multirow{-2}{*}{63}
& \multirow{-2}{*}{\textbf{0}}
& 0.885
& \multirow{-2}{*}{55208}
& \multirow{-2}{*}{0.995}
& \multirow{-2}{*}{182}
& \multirow{-2}{*}{\textbf{0}}
& \textbf{0.919}
& \multirow{-2}{*}{50692}
& \multirow{-2}{*}{0.992}
& \multirow{-2}{*}{893} \\
\hline

& & 0.983
& & 0.967 & & & 
& & 0.977 & & & 
& & 0.965 & & &
& & \textbf{0.979} & & & \\
\multirow{-2}{*}{WDM}
& \multirow{-2}{*}{{\color[HTML]{C00000}594}}
& 0.983
& \multirow{-2}{*}{{\color[HTML]{C00000}533}}
& 0.963
& \multirow{-2}{*}{70264}
& \multirow{-2}{*}{0.997}
& \multirow{-2}{*}{55}
& \multirow{-2}{*}{{\color[HTML]{C00000}430}}
& {\color[HTML]{808080}0.980}
& \multirow{-2}{*}{136818}
& \multirow{-2}{*}{0.987}
& \multirow{-2}{*}{139}
& \multirow{-2}{*}{\textbf{0}}
& 0.961
& \multirow{-2}{*}{71829}
& \multirow{-2}{*}{0.997}
& \multirow{-2}{*}{288}
& \multirow{-2}{*}{\textbf{0}}
& \textbf{0.976}
& \multirow{-2}{*}{70924}
& \multirow{-2}{*}{0.992}
& \multirow{-2}{*}{2268} \\
\hline

& & 0.835
& & 0.817 & & & 
& & 0.833 & & & 
& & 0.822 & & &
& & \textbf{0.833} & & & \\
\multirow{-2}{*}{TTS}
& \multirow{-2}{*}{{\color[HTML]{C00000}799}}
& 0.859
& \multirow{-2}{*}{{\color[HTML]{C00000}624}}
& 0.773
& \multirow{-2}{*}{88285}
& \multirow{-2}{*}{0.994}
& \multirow{-2}{*}{56}
& \multirow{-2}{*}{{\color[HTML]{C00000}630}}
& {\color[HTML]{808080}0.847}
& \multirow{-2}{*}{136001}
& \multirow{-2}{*}{0.986}
& \multirow{-2}{*}{139}
& \multirow{-2}{*}{\textbf{0}}
& 0.768
& \multirow{-2}{*}{89869}
& \multirow{-2}{*}{0.995}
& \multirow{-2}{*}{330}
& \multirow{-2}{*}{\textbf{0}}
& \textbf{0.842}
& \multirow{-2}{*}{81395}
& \multirow{-2}{*}{0.988}
& \multirow{-2}{*}{2260} \\
\hline

\end{tabular}
}
\label{tab:main_result}
\end{table*}

%% file: figtex/fig_fixing_comparison.tex
\begin{figure}
    \centering
    \includegraphics[width=0.95\columnwidth]{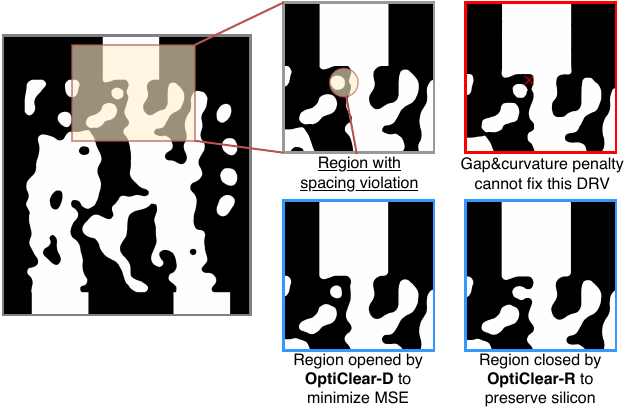}
    \vspace{-5pt}
    \caption{A spacing violation in a performance-sensitive region near the input port. This DRV cannot be removed by gap or curvature penalties alone. To minimize geometry displacement, \nameD opens the violating region, while \nameR closes the region to preserve the silicon connectivity.}
    \label{fig:fixing_comparison}
     \vspace{-10pt}
\end{figure}

%% file: figtex/fig_ablation_mse.tex
\begin{figure}
    \centering
    \includegraphics[width=1\columnwidth]{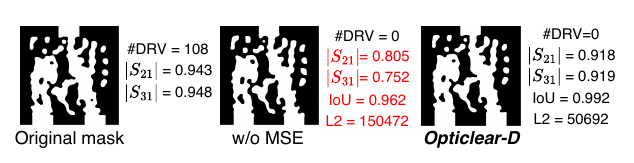}
    \vspace{-10pt}
    \caption{The MSE loss in \nameD suppresses mask distortion with better function preservation during legalization.}
    \label{fig:MSE_compare}
     \vspace{-5pt}
\end{figure}

%% file: tables/tab_exp_mfs.tex
\begin{table*}[t]
\caption{Evaluation of figure-of-merit (FoM) ($|S_{21}|,|S_{12}|$ for OD, $|S_{21}|,|S_{31}|$ for MDM/WDM), \#DRV among original mask, \nameR, and \nameD on three devices designed with different minimum feature size (MFS) (nm).
}
\vspace{-8pt}
\resizebox{0.99\textwidth}{!}{
\begin{tabular}{|c|cccccc|cccccc|cccccc|}
\hline
\multirow{3}{*}{MFS} & \multicolumn{6}{c|}{Optical diode (OD)}                                                                & \multicolumn{6}{c|}{Mode-division multiplexer (MDM)}                                                 & \multicolumn{6}{c|}{Wavelength-division   multiplexer (WDM)}                                           \\
                     & \multicolumn{2}{c}{Design mask} & \multicolumn{2}{c}{\nameR} & \multicolumn{2}{c|}{\nameD} & \multicolumn{2}{c}{Design mask} & \multicolumn{2}{c}{\nameR} & \multicolumn{2}{c|}{\nameD} & \multicolumn{2}{c}{Design mask} & \multicolumn{2}{c}{\nameR} & \multicolumn{2}{c|}{\nameD} \\
                     & FoM   & \#DRV                & FoM    & \#DRV              & FoM    & \#DRV             & FoM   & \#DRV                & FoM   & \#DRV               & FoM   & \#DRV              & FoM   & \#DRV                & FoM   & \#DRV               & FoM   & \#DRV               \\\hline
\multirow{2}{*}{70}  & 0.997   & \multirow{2}{*}{\color[HTML]{C00000}536}  & 0.892    & \multirow{2}{*}{0}  & 0.955    & \multirow{2}{*}{0} & 0.942   & \multirow{2}{*}{\color[HTML]{C00000}315}  & 0.889   & \multirow{2}{*}{0}   & 0.912   & \multirow{2}{*}{0}  & 0.984   & \multirow{2}{*}{\color[HTML]{C00000}916}  & 0.946   & \multirow{2}{*}{0}  & 0.981   & \multirow{2}{*}{0}  \\
                     & 6.2E-5 &                       & 2.5E-5  &                     & 5.0E-4  &                    & 0.961   &                       & 0.826   &                      & 0.907   &                     & 0.983   &                       & 0.969   &                      & 0.981   &                      \\\hline
\multirow{2}{*}{90}  & 0.965   & \multirow{2}{*}{\color[HTML]{C00000}2235} & 0.965    & \multirow{2}{*}{0}  & 0.962    & \multirow{2}{*}{0} & 0.941   & \multirow{2}{*}{\color[HTML]{C00000}266}  & 0.908   & \multirow{2}{*}{0}   & 0.909   & \multirow{2}{*}{0}  & 0.981   & \multirow{2}{*}{\color[HTML]{C00000}708}  & 0.975   & \multirow{2}{*}{0}   & 0.979   & \multirow{2}{*}{0}   \\
                     & 4.8E-3 &                       & 2.5E-3  &                     & 5.2E-3  &                    & 0.924   &                       & 0.888   &                      & 0.887   &                     & 0.983   &                       & 0.966   &                      & 0.977   &                      \\\hline
\multirow{2}{*}{110} & 0.997   & \multirow{2}{*}{\color[HTML]{C00000}407}  & 0.997    & \multirow{2}{*}{0}  & 0.997    & \multirow{2}{*}{0} & 0.905   & \multirow{2}{*}{\color[HTML]{C00000}193}  & 0.905   & \multirow{2}{*}{0}   & 0.905   & \multirow{2}{*}{0}  & 0.982   & \multirow{2}{*}{\color[HTML]{C00000}671}  & 0.974   & \multirow{2}{*}{0}   & 0.974   & \multirow{2}{*}{0}   \\
                     & 2.3E-5 &                       & 3.1E-3  &                     & 2.1E-5  &                    & 0.923   &                       & 0.918   &                      & 0.917   &                     & 0.982   &                       & 0.965   &                      & 0.962   &                      \\\hline
\multirow{2}{*}{130} & 0.978   & \multirow{2}{*}{\color[HTML]{C00000}400}  & 0.978    & \multirow{2}{*}{0}  & 0.978    & \multirow{2}{*}{0} & 0.934   & \multirow{2}{*}{\color[HTML]{C00000}273}  & 0.933   & \multirow{2}{*}{0}   & 0.932   & \multirow{2}{*}{0}  & 0.967   & \multirow{2}{*}{\color[HTML]{C00000}87}   & 0.965   & \multirow{2}{*}{0}   & 0.965   & \multirow{2}{*}{0}   \\
                     & 3.2E-3 &                       & 1.6E-3  &                     & 3.2E-3  &                    & 0.915   &                       & 0.916   &                      & 0.914   &                     & 0.968   &                       & 0.966   &                      & 0.966   &                      \\\hline
\multirow{2}{*}{150} & 0.967   & \multirow{2}{*}{\color[HTML]{C00000}11}   & 0.968    & \multirow{2}{*}{0}  & 0.968    & \multirow{2}{*}{0} & 0.902   & \multirow{2}{*}{\color[HTML]{C00000}54}   & 0.901   & \multirow{2}{*}{0}   & 0.902   & \multirow{2}{*}{0}  & 0.976   & \multirow{2}{*}{\color[HTML]{C00000}272}  & 0.975   & \multirow{2}{*}{0}   & 0.974   & \multirow{2}{*}{0}   \\
                     & 3.2E-3 &                       & 4.8E-3  &                     & 3.2E-3  &                    & 0.877   &                       & 0.876   &                      & 0.877   &                     & 0.968   &                       & 0.967   &                      & 0.967   &                      \\ \hline
\end{tabular}
}
\label{tab:mfs_result}
\end{table*}

%% file: tables/tab_exp_design_rule.tex
\begin{table*}[t]
\caption{Evaluation of \nameR and \nameD on three devices (designed with MFS=110 nm) across different spacing/width rules (nm) in terms of \#DRV, FoM ($|S_{21}|,|S_{12}|$ for OD, $|S_{21}|,|S_{31}|$ for MDM/WDM), and runtime (s).}
\vspace{-8pt}
\resizebox{0.99\textwidth}{!}{
\begin{tabular}{|cc|cccc|cccc|cccc|}
\hline
\multirow{2}{*}{Method}     & \multirow{2}{*}{Metrics}      & \multicolumn{4}{c|}{OD: $|S_{21}|=0.997$,   $|S_{12}|=$2.3E-5} & \multicolumn{4}{c|}{MDM: $|S_{21}|=0.905$,   $|S_{31}|=0.923$} & \multicolumn{4}{c|}{WDM: $|S_{21}|=0.982$,   $|S_{31}|=0.982$} \\
                            &  & Rule: 70      & Rule:  90      & Rule: 110     & Rule: 130     & Rule: 70      & Rule:  90     & Rule: 110     & Rule: 130     & Rule: 70      & Rule:  90      & Rule: 110     & Rule: 130     \\\hline
\multirow{4}{*}{\nameR} & \#DRV                           & 0             & 0              & 0            & 0            & 0             & 0             & 0             & 0             & 0             & 0              & 0            & 0            \\
                             & $|S_{21}|$               & 0.997         & 0.991          & 0.985         & 0.762         & 0.905         & 0.876         & 0.871         & 0.847         & 0.974         & 0.920          & 0.902         & 0.838         \\
                                                   & $|S_{31}|$ or $|S_{12}|$ & 3.1E-3       & 2.0E-4        & 1.3E-3       & 3.3E-3       & 0.918         & 0.847         & 0.849         & 0.522         & 0.965         & 0.953          & 0.874         & 0.418         \\
                            & Runtime (s)                            & 150           & 151            & 552           & 903           & 152           & 232           & 238           & 309           & 302           & 567            & 596           & 712           \\\hline
\multirow{4}{*}{\nameD}  & \#DRV                           & 0             & 0              & 0            & 0            & 0             & 0             & 0            & 0            & 0             & 0              & 0            & 0            \\
                            & $|S_{21}|$                      & 0.997         & 0.986          & 0.990         & 0.984         & 0.905         & 0.885         & 0.887         & 0.884         & 0.974         & 0.966          & 0.951         & 0.885         \\
                                                    & $|S_{31}|$ or $|S_{12}|$ & 2.1E-5       & 6.0E-4        & 4.0E-4       & 6.0E-4       & 0.917         & 0.887         & 0.859         & 0.890         & 0.962         & 0.913          & 0.890         & 0.830         \\
                            & Runtime (s)                           & 1197          & 1268           & 1216          & 1271          & 886           & 887           & 889           & 892           & 2257          & 2716           & 2253          & 2264          \\ \hline
\end{tabular}
}
\label{tab:design_rule_result}
\end{table*}

%% file: tables/tab_exp_resolution.tex
\begin{table}
\centering
\vspace{-5pt}
\caption{Legalization results of the MDM device with different mask resolutions using \nameD.}
\vspace{-5pt}
\resizebox{\columnwidth}{!}{
\begin{tabular}{|c|cllccc|}
\hline
Mask   resolution & \#DRV & \multicolumn{1}{c}{$|S_{21}|$} & \multicolumn{1}{c}{$|S_{31}|$} & L2    & IoU   & RT (s) \\ \hline
4 nm/pixel               & 90    & 0.921                          & 0.921                          & 62617 & 0.982 & 61     \\ \hline
2 nm/pixel                  & 78    & 0.913                          & 0.899                          & 55957 & 0.986 & 111    \\ \hline
1 nm/pixel           & 0     & 0.918                          & 0.919                          & 50692 & 0.992 & 893    \\ \hline
\end{tabular}
}
\label{tab:exp_resolution}
\end{table}

%% file: doc/5_conclu.tex
\section{Conclusion}
\label{sec:Conclusion}
We presented \name, the first curvilinear design-rule legalization framework for inverse-designed photonic devices. 
It offers two legalization engines: \nameR for efficient rule-based violation resolution, and \nameD for differentiable minimum-distortion legalization under morphological stationary-point constraints.
We further develop customized differentiable morphological GPU operators that substantially improve scalability for legalizing high-resolution masks.
Extensive evaluation across diverse inverse-designed devices and rule settings shows that \name reduces DRVs from thousands to zero while preserving the pre-legalization geometry and optical functionality, with the rule-based legalizer providing high efficiency and the differentiable legalizer achieving better post-legalization fidelity.
Beyond its immediate role in post-design legalization, this work lays a foundation for EPDA support of curvilinear photonic layouts and opens a path toward differentiable inverse-design flows with explicit design-rule awareness.